\documentclass[epj]{svjour}

\usepackage{graphicx}
\usepackage{dcolumn}
\usepackage{bm}
\usepackage{subfigure}
\usepackage{amsmath}
\usepackage{exscale}

\begin{document}

\title{Fluctuating semiflexible polymer ribbon constrained to a ring}

 \author{Karen Alim \and Erwin Frey}
\institute{Arnold Sommerfeld Center for Theoretical Physics and Center for NanoScience,
Department of Physics,\\ Ludwig-Maximilians-Universit\"at M\"unchen,
Theresienstrasse 37, D-80333 M\"unchen, Germany
}

\date{9 November 2007}

\abstract{
Twist stiffness and an asymmetric bending stiffness of a polymer or a polymer bundle is captured by the elastic ribbon model. We investigate the effects a ring geometry induces to a thermally fluctuating ribbon, finding bend-bend coupling in addition to twist-bend coupling. Furthermore, due to the geometric constraint the polymer's effective bending stiffness increases. A new parameter for experimental investigations of polymer bundles is proposed: the mean square diameter of a ribbonlike ring, which is determined analytically in the semiflexible limit. Monte Carlo simulations are performed which affirm the model's prediction up to high flexibility.
\PACS{{87.15.Aa}{Theory and modeling, computer simulation}\and
  {87.15.Ya}{Fluctuations}\and  
  {36.20.Ey}{Conformation (statics and dynamics)\and
  {05.40.-a}{Fluctuation phenomena, random processes, noise, and Brownian motion}
}}
}

\maketitle
\section{Introduction}
\label{sec_introduction}
The semiflexible polymers constituting the cell's cytoskeleton determine morphology and elasticity of the cell. By use of binding proteins the polymers organize into bundles of closely packed filaments, which form the building blocks for a variety of cellular processes such as filopodia, villi or cilia \cite{adams04,bartles00,tilney89,lin94}.
Recent investigations of the mechanical properties of polymer bundles \cite{claessens} pose the question for a coarse-grained description of their mechanics beyond the macroscopic models for semiflexible polymers such as the Kratky-Porod model \cite{kratkyporod} and its continuous description \cite{saito}, which represent the polymer by a space curve with a single material parameter. A more general description is obtained by the classical elastic ribbon \cite{landauelas}, which accounts for three parameters concerning  bending and  twisting. 
There are also advances to describe polymer bundles as a set of interconnected semiflexible polymers, starting out with railway track models of two coupled filaments in two \cite{everaers} and three \cite{liverpool,golestanian,mergell} dimensions, whose exact implementation and results were much in debate \cite{mergell}. Recently, a model for wormlike bundles with isotropic cross-section, consisting of a finite number of semiflexible polymers, has been introduced and analyzed \cite{heussinger,bathe}. These models show new material properties, such as length-dependent bending and twisting moduli \cite{heussinger}, as they include stretching and shearing terms, not accounted for in the elastic ribbon model considered in this paper. The latter regards both bending and twist moduli as fixed material parameters.             

Nature not only imposes geometric constraints on bio\-poly\-mers by confinement through membranes but also deliberately uses the advantages of certain geometries as in the case of circular DNA. Especially the ring geometry induces interesting effects. In the context of plectonemes mechanical equilibria of elastic rings have been studied extensively \cite{benhamdna}, whereas thermal equilibria have mainly been addressed for flexible rings \cite{shimada,hearst,klenin}. Only recently investigations dealt with the thermal fluctuations of polymer ribbons that are stress-free in a circular configuration \cite{rabin01}. However, both the mechanical and the thermal motion of rings with intrinsic bend differ principally from initially straight polymers as has been observed for the plectonem transition \cite{bauer}. For biopolymers participating in the cytoskeleton such as actin and microtubuli the stress-free conformation is straight. Even small DNA rings have lately been found to exhibit no intrinsic bending without attachment of proteins \cite{cloutier}. 
Furthermore, ribbonlike rings require theoretical modeling as in vitro experiments have shown that the mechanical properties of polymer bundles can be well studied when the bundle is constrained to a ring structure \cite{claessens}. In vivo polymer bundle rings are found in erythrocytes of birds and reptiles \cite{elbaum}.

Ligating a ribbon's end to form a ring reduces the set of conformations available to the ribbon. Perturbing the ring's mechanical equilibrium configuration by a small twist induces a bending of the ribbon's center line, as can be easily visualized with a ring consisting of a small strip of paper. This twist-bend coupling is mathematically captured by White's formula \cite{white,fuller}, which connects the overall twist to the global configurational integral denoted writhe for any closed curve. However, we will show that for a ribbonlike ring with asymmetric cross section bending within the equilibrium plane and bending transversal to the equilibrium plane are also coupled. In this case it is instructive to parameterize the ring in terms of Euler angles such that each of them describes either a bending or a twisting motion. Modeling thermally fluctuating rings we pursue to give analytical expressions for experimentally observable properties of the ring. In analogy to the mean square end-to-end distance of a polymer the mean square diameter is well suited to represent the statistical properties of a fluctuating ribbonlike ring.

In this work we investigate the effects of a ring geo\-me\-try on a fluctuating semiflexible polymer ribbon. In Section \ref{sec_model} the elastic ribbon model is formulated in the limit of small fluctuations about a ring. The elastic free energy already indicates the first effect of the ring geometry on a ribbon: the bend-bend coupling in addition to twist-bend coupling. Furthermore, in Section \ref{sec_model} the foundations are laid for the analytic calculation of the experimentally accessible mean square diameter. The discussion of this ensemble parameter in Section \ref{sec_results} reveals the quality of the coupling and the second effect due to the geometric constraint: the effective stiffening of the ribbon. Assessing the quality of the analytical result for the mean square dia\-me\-ter with Monte Carlo simulation we find that our model predicts the correct behavior for a polymer with symmetric cross section up to total contour length seven times the persistence length. Beyond this stiff limit our simulations confirm previous results for almost flexible polymer rings. Thus, polymer rings with symmetric cross section are fully characterized analytically.

\section{The model}
\label{sec_model}
To specify a deformed state of a ribbon a local body coordinate system $\{\mathbf{t}_1(s)$, $\mathbf{t}_2(s)$, $\mathbf{t}_3(s)\}$ is assigned to every point $s\in[0,L]$ along the center line of the ribbon, such that the $\mathbf{t}_3(s)$ axis points tangent to the center line of the rod in the direction of increasing $s$, while $\mathbf{t}_1(s)$ and $\mathbf{t}_2(s)$ are aligned to the principal axes of the cross section. The change of body coordinates along the inextensible ribbon is described by rotations $\boldsymbol{\omega}$ via the generalized Frenet equations:
\begin{equation}
\frac{d\mathbf{t}_i}{ds}=\boldsymbol{\omega}\times\mathbf{t}_i,\: i=1,2,3.
\end{equation}
$\omega _1$ and $\omega_2$ may be interpreted as the curvatures in the principal directions, while $\omega_3$ may be denoted as the helical deformation density~\cite{benhamdna}. Hence, the elastic free energy of a ribbon reads \cite{landauelas}:
 \begin{equation}
F=\frac{\mathit{k_{B}}T}{2}\int_{0}^{L}ds\left[a_{1}\omega_{1}^2+a_{2}\omega_{2}^2+a_{3}\omega_{3}^2\right],
\label{landauelasticfree}
\end{equation} 
where $a_1$ and $a_2$ denote the bending stiffnesses along the principal axes of the cross section and $a_3$ indicates the twist stiffness. For the geometry of a ring it is instructive to express the rotations $\boldsymbol{\omega}$ in terms of Euler angles $\phi(s), \theta(s),\psi(s)$ according to Euler's equations, where $\phi$ is the azimuthal angle, $\theta$ the polar angle and $\psi$ the twist angle. The constraint of a closed ring of contour length $L=2\pi R_c$ is then satisfied by $\phi_0(s)=s/R_c$ and $\theta_0(s)=\pi/2$. The remaining angle $\psi_0(s)$ is chosen as to minimize the elastic free energy which is achieved by $\psi_0(s)=\pi/2$ for $a_1<a_2$ or by $\psi_0(s)=0$ for $a_1>a_2$. These two cases are essentially the same, as the major principal axis of the cross section is always perpendicular to the equilibrium plane of the ring. We therefore restrict our discussion to  $a_1<a_2$. As the biopolymers constituting polymer bundles are itself semiflexible, we investigate the limit of small fluctuations of the ribbonlike ring. In this limit self-avoidance is satisfied and hence our phantom chain model should yield correct results for real polymers. Furthermore, we keep the total length of the ribbon strictly fixed. Small fluctuations in the contour length might though lead to interesting behavior, the investigation of which lies beyond the scope of this publication. 
\begin{figure}[t]
\centering
\includegraphics[width=0.45\textwidth]{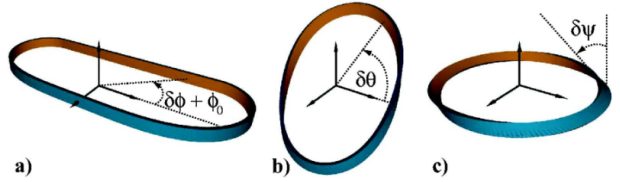}
\caption{Illustration of the Euler angles and their coupling induced by the ring geometry: A sinusoidal variation of each Euler angle at a time results in more than one free energy term, if coupling is present: (a) $\delta\phi=\sin(s/R_c)$, in-plane bending only, $\tilde{F}=a_1$, (b) $\delta\theta=\sin(s/R_c)$, transversal bending yields in-plane bending and twisting, $\tilde{F}=-a_1+a_2+a_3$, (c) $\delta\psi=\sin(s/R_c)$, twisting yields both in-plane and transversal bending, $\tilde{F}=-a_1+a_2+a_3$; $F$ in units of $\pi \mathit{k_B} T/2R_c$.}
\label{fig_examples}
\end{figure}\\

Assuming small deviations $\delta\phi,\delta\theta,\delta\psi$ from the ground state of the ring specified by $\phi_0,\theta_0,\psi_0$ the rotations $\boldsymbol\omega$ are expanded and all quadratic terms obtained from the squared rotations are taken into account to obtain the elastic free energy in the limit of small fluctuations: 
\begin{multline}
\lefteqn{F=\frac{\mathit{k_{B}}T}{2}\int_{0}^{L}ds \Bigg\{a_{1}\left[\left(\frac{d \delta \phi}{ds}\right)^2-\left(\frac{ \delta \theta}{R_c}\right)^2+\left(\frac{d\delta \theta}{ds}\right)^2\right]}\\
+(a_{2}-a_{1})\left(\frac{\delta \psi}{R_c}+\frac{d\delta \theta}{ds}\right)^2
+a_{3}\left(\frac{ d\delta \psi}{ds}-\frac{\delta \theta}{R_c}\right)^2\Bigg\}.
\label{eq_freeelastic}
\end{multline}
In parameterizing the ribbon in terms of Euler angles a distinct motion was assigned to each angle. The azimuthal angle $\delta\phi$ describes bending within the equilibrium plane, while the polar angle $\delta\theta$ characterizes transversal bending and the twist angle $\delta\psi$ depicts twisting around the center line of the ribbon, see Fig.~\ref{fig_examples}. Considering the elastic free energy associated with each conformation of the ribbon, three energy components can be distinguished, the in-plane bending energy $F_{\mathrm{in-plane}}$, the transversal bending energy $F_{\mathrm{trans}}$ and the twisting energy $F_{\mathrm{twist}}$, recognizable by their proportionality to their associated stiffnesses $a_1$, $a_2$ and $a_3$, respectively. In a completely decoupled system such as an open polymer ribbon each energy component would comprise just the spatial derivative of its corresponding Euler angle
\begin{multline}
F=\frac{\mathit{k_B}T}{2}\int_0^L ds \Bigg\{a_1\left(\frac{d\delta\phi}{ds}\right)^2+a_2\left(\frac{d\delta\theta}{ds}\right)^2\\+a_3\left(\frac{d\delta\psi}{ds}\right)^2\Bigg\}\,.
\end{multline} 
In the limit of $R_c\to \infty$, keeping $a_1$, $a_2$ and $a_3$ fixed, the elastic free energy of a polymer ribbon ring as given in equation (\ref{eq_freeelastic}) decouples to the above expression, as local fluctuations are not influenced by the ring geometry anymore. However, for small contour radius with respect to the bending and twisting stiffnesses, the elastic free energy of a polymer ribbon ring exhibits both twist-bend coupling and bend-bend coupling. Twist-bend coupling heuristically means that twisting induces bending and vice versa. In the case of $F_{\mathrm{twist}}$ the twist arising form the change of the twist angle $\delta\psi$ is diminished by the height $\delta\theta$ relative to the contour radius $R_c$ the ring gains through transversal bending. The opposite coupling effect is observed for $F_{\mathrm{trans}}$ where the bending energy is augmented by the local twist relative to $R_c$. Analyzing the in-plane bending energy component 
\begin{multline}
F_{\mathrm{in-plane}}=\frac{\mathit{k_B}T}{2}\int_0^L ds\Bigg\{\left(\frac{d\delta\phi}{ds}\right)^2-\left(\frac{\delta\theta}{R_c}\right)^2\\-\left(\frac{\delta\psi}{R_c}\right)^2-2\frac{d\delta\theta}{ds}\frac{\delta\psi}{R_c}\Bigg\}\,,
\end{multline}
 twist-bend coupling is found again, but now also a variation of the transversal bending Euler angle $\delta\theta$ reduces the in-plane bending energy. This is identified as bend-bend coupling. Furthermore, the release by the twist depends on the curvature of the transversal bending, hence, a mixed term arises.

As a demonstrative example the conformations of a ribbon with sinusoidal variation of each Euler angle are depicted in Fig.~\ref{fig_examples}. Both altering $\delta\theta$ and $\delta\psi$ causes contributions to all three terms in the elastic free energy exemplifying the couplings. Remarkably, altering the twist angle $\delta \psi$ or the polar angle $\delta\theta$ in the same sinusoidal manner requires exactly the same energy cost. Furthermore, for a ribbon with symmetric cross section the only energy contribution due to sinusoidal transversal bending or twisting is proportional to the twist stiffness only, as other terms cancel. Previous work assumed the unstressed state of a polymer to be a ring conformation \cite{rabin01}, whose applicability to cytoskeletal polymers maybe limited, yielding no coupling of transversal bending and twisting to in-plane bending. 
\label{sec_msd}\\

Given the elastic free energy statistical properties may be derived. Because of its importance for experimental investigations we shall focus on the mean square diameter of the ring. Since the Euler angles must obey periodic boundary conditions to satisfy a closed ring they are expanded in a Fourier series. In the Fourier representation the periodic boundary condition correspond to: $\delta\tilde{\theta}(0)=\delta\tilde{\phi}(1)=0$. The Fourier transformed elastic free energy then results in:
\begin{multline}
\lefteqn{F=\frac{\mathit{k_{B}}T\pi}{R_c}\Big\{(a_2-a_1)|\delta\tilde{\psi}(0)|^2}\\
+(a_3+a_2-a_1)|\mathrm{i} \delta\tilde{ \theta}(1)+\delta\tilde{ \psi}(1)|^2\\
+\sum_{n=2}^{\infty}\big[(a_2-a_1)|\mathrm{i} n \delta\tilde{ \theta}(n)+\delta\tilde{ \psi}(n)|^2+a_1n^2|\delta\tilde{ \phi}(n)|^2\\
+a_1(n^2-1)|\delta\tilde{ \theta}(n)|^2+a_3|\mathrm{i} n\delta\tilde{ \psi}(n)-\delta\tilde{ \theta}(n)|^2\big]\Big\}\:.
\end{multline}
Note that concerning the Euler angles themselves only the twist angle $\delta\psi$ and the polar angle $\delta\theta$ are coupled. Diagonalizing the Fourier transformed elastic free energy and applying the equipartition theorem the correlations of the Euler angles are obtained depending on the mean square modes.
\begin{eqnarray}
\langle\delta \phi(s_2)\delta \phi(s_1)\rangle&=&\frac{R_c}{\pi}\sum_{n=2}^{\infty}\langle\delta \tilde{\phi}^2(n)\rangle\cos(ns/R_c)\;,\label{eq_phiphi}\\
\langle\delta \theta(s_2)\delta \theta(s_1)\rangle&=&\frac{R_c}{\pi}\frac{\cos(s/R_c)}{a_3+a_2-a_1}\nonumber\\
&&+\frac{R_c}{\pi}\sum_{n=2}^{\infty}\langle\delta \tilde{\theta}^2(n)\rangle\cos(ns/R_c)\;,\label{eq_thetatheta}\\
\langle\delta \psi(s_2)\delta \psi(s_1)\rangle&=&\frac{R_c}{\pi(a_2-a_1)}+\frac{R_c}{\pi}\frac{\cos(s/R_c)}{a_3+a_2-a_1}\nonumber\\
&&+\frac{R_c}{\pi}\sum_{n=2}^{\infty} \langle\delta \tilde{\psi}^2(n)\rangle\cos(ns/R_c)\;,\label{eq_psipsi}
\end{eqnarray}
where $s=|s_2-s_1|$. Closed expressions for the modes are only obtained for $n\geq2$. For the azimuthal angle $\delta \phi$ the zeroth mode vanishes due to the rotational symmetry around the axis through the ring's center of mass. The first mode is zero due to the periodic boundary conditions, which also set the zeroth polar angle mode equal nought. However, the zeroth mode of the twist angle $\delta\psi$ vanishes only in the limit of $(a_2-a_1)\to \infty$. The minimum of the elastic free energy is given by $\delta \psi=0$, such that the major principal axis of the cross section along $a_2$ in our definition is perpendicular to the equilibrium plane. For finite values of both bending stiffnesses small fluctuations around $\delta\psi=0$ arise with an amplitude that increases with a decaying difference between the bending stiffnesses, being not defined for equal magnitude.  Among all three stiffnesses $a_1$ influences the fluctuations of a polymer ribbon ring the most. For $a_1=\infty$ all deformations of the ring are energetically hindered, since also twisting and transversal bending require suppressed in-plane bending due to the coupling. For small $a_1$ on the other hand the fluctuations exceed the small deviations approximation. As the effect of the other two parameter is smaller it is reasonable to study their significance relative to the in-plane bending stiffness $a_1$. Hence, we introduce the dimensionless parameters $\alpha=a_1/a_2$ and $\tau=a_1/a_3$, in which the modes of the Euler angles for $n\geq2$ are given by:    
\begin{eqnarray}
\langle\delta \tilde{\phi}^2(n)\rangle&=&\frac{R_c}{a_{1} n^2}\;,\label{eq_modephi}\\
\langle\delta \tilde{\theta}^2(n)\rangle&=&\frac{R_c}{a_1}\left[\frac{1}{n^2-1}+\frac{\alpha-1}{n^2+(1-\tau)(\alpha-1)}\right],\label{eq_modetheta}\\
\langle\delta \tilde{\psi}^2(n)\rangle&=&\frac{R_c}{a_1}\left[\frac{1}{n^2-1}+\frac{\tau-1}{n^2+(1-\tau)(\alpha-1)}\right].\label{eq_modepsi}
\end{eqnarray}
As variations in $\delta\phi$ do not influence transversal bending or twisting, the modes only depend on the in-plane bending stiffness $a_1$, whereas the modes of the multiply coupled Euler angles $\delta\theta$ and $\delta\phi$ are functions of all three stiffnesses.
 
 The mean square diameter is by definition just the mean square distance between two positions $\mathbf{r}(s)$ separated by half the contour length $L/2=\pi R_c$ along the center line of the ribbon. The mean square distance itself can be calculated from the tangent-tangent correlation by: 
\begin{multline}
\langle D^2\rangle=\langle\left(\mathbf{r}(L/2)-\mathbf{r}(0)\right)^2\rangle\\
=\int_0^{\frac{L}{2}} dy\int_{0}^{\frac{L}{2}} dy'\langle \vec{t}_3(y)\vec{t}_3(y')\rangle\:.
\label{rsquared}
\end{multline}
The tangent is approximated for small deviations from the rigid ring up to second order in $\delta\phi$ and $\delta\theta$ such that the tangent-tangent correlation yields: 
\begin{multline}
\langle \vec{t}_3(s_1)\vec{t}_3(s_2)\rangle=\langle\delta \phi(s_1)\delta \phi(s_2)\rangle\cos\bigg(\frac{|s_1-s_2|}{R_c}\bigg)\\
+\langle\delta \theta(s_1)\delta \theta(s_2)\rangle\\
+\left[1-\langle\delta \theta^2\rangle-\langle\delta \phi^2\rangle\right]\cos\bigg(\frac{|s_1-s_2|}{R_c}\bigg).
\label{tangenttangent}
\end{multline} 
Evaluating the double integrals the mean square diameter is represented in terms of the modes of the Euler angles:
\begin{eqnarray}
\langle D^2\rangle&=&(2R_c)^2-\frac{4R_c^2}{\pi}\sum_{n=2}^{\infty}\langle\delta\tilde{\phi}^2(n)\rangle-\frac{4R_c^2}{\pi}\sum_{n=2}^{\infty}\langle\delta\tilde{\theta}^2(n)\rangle\nonumber\\
&+&\frac{4R_c^2}{\pi}\sum_{\begin{subarray}{l}n=2\\ n\;\mathrm{even}\end{subarray}}^{\infty}\frac{(n^2+1)\langle\delta\tilde{\phi}^2(n)\rangle}{(n^2-1)^2}\nonumber\\
&+&\frac{4R_c^2}{\pi}\sum_{\begin{subarray}{l}n=3\\ n\;\mathrm{odd}\end{subarray}}^{\infty}\frac{\langle\delta\tilde{\theta}^2(n)\rangle}{n^2}.
\label{eq_d2}
\end{eqnarray}
From this expression the mean square diameter can be derived. The sum of all four sums always yields a negative term corresponding to a decrease of the mean squared diameter from its rigid ring value of $D=2R_c$.
Note that the mean square diameter does not depend on the twist modes since twist motions in the weakly fluctuating approximation do not affect the center line of the ribbon and hence do not decrease the mean square diameter.
\section{Results and Discussion}
\label{sec_results}
To study the effect of a ring geometry on a fluctuating polymer ribbon with minor bending stiffness $a_1$, relative bending stiffness $\alpha=a_1/a_2$ and relative twist stiffness $\tau=a_1/a_3$ we choose the mean square diameter as parameter for the ensemble average since this parameter is experimentally accessible and enables comparison with the mean squared end-to-end distance of an open polymer. In the limit of small fluctuations the analysis in the previous section enables us to evaluate the mean square diameter of a polymer ribbon constrained to a ring by inserting the modes of the Euler angles as given in equations (\ref{eq_modephi}) and (\ref{eq_modetheta}) into the formula (\ref{eq_d2}) and calculating the sums, resulting in:
\begin{multline}
\langle D^2\rangle=4R_c^2-\frac{R_c^3}{a_1}\bigg[\frac{4+\pi^2\tau}{2\pi(\tau-1)}-\frac{\tan(\pi\sqrt{(1-\alpha)(1-\tau)}/2)}{\sqrt{(1-\alpha)(1-\tau)^3}}\\
-2\sqrt{\frac{1-\alpha}{1-\tau}}\cot(\pi\sqrt{(1-\alpha)(1-\tau)})\bigg] \;. 
\label{eq_diameter}
\end{multline}
Analyzing this mean value the effects of the ring geometry are unveiled, namely coupling of twisting and bending and an effective stiffening of the polymer. 
\begin{figure}[t]
\centering
\includegraphics[width=0.35\textwidth]{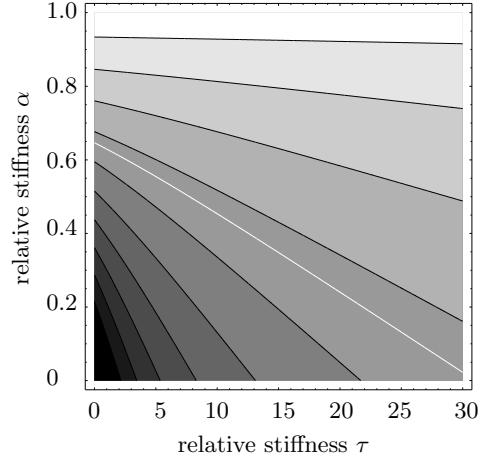}
\caption{ Contour plot of the descent of the mean square diameter $ -\partial\langle D^2\rangle/\partial (R_c^3/a_1)$ versus $\alpha=a_1/a_2$ and $\tau=a_1/a_3$, where the contour lines decrease from $\pi/2$ at $(\infty,1)$ to $2/\pi$ at $(0,0)$ in ten equal steps. The additional white contour marks where in-plane and transversal fluctuations are equal in decreasing the diameter.}
\label{fig_diametergen}
\end{figure}
To this end a plot of the diameter is desirable. Since that would require a four dimensional space, we refrain from depicting the reciprocally proportionally decrease of the mean square diameter with bending stiffness $a_1$ and show a contour plot of the descent of the mean square diameter  $ -\partial\langle D^2\rangle/\partial (R_c^3/a_1)$ versus $\alpha$ and $\tau$ in Fig.~\ref{fig_diametergen}. The diameter decreases by  $2/\pi \cdot R_c^3/a_1$ at $(\tau=0,\alpha=0)$ up to $\pi/2\cdot R_c^3/a_1$ at $(\infty,1)$. 

Our analysis of the elastic free energy in the limit of small fluctuations shows a bend-bend coupling in addition to twist-bend coupling. The motion describing twisting $\delta\psi$ in equation (\ref{eq_freeelastic}) contributes to the transversal and the in-plane bending term in the elastic free energy. Vice versa the motion characterizing transversal bending $\delta\theta$ increases the twisting free energy term, hence bending and twisting are coupled. Furthermore, transversal bending motion indicated by $\delta\theta$ in equation (\ref{eq_freeelastic}) decrease the in-plane bending term in the elastic free energy, therefore both bending motions are coupled. In the case of the mean square diameter the coupling comes apparent in the complicated dependence on the bending and twisting stiffness in equation (\ref{eq_diameter}), in contrast to the uncoupled case, where the subtrahend of the diameter is constituted of three independent term proportional to the inverse of the bending or twisting stiffness, respectively. Both transversal bending and twisting are strongly dependent on the in-plane bending stiffness, as all fluctuations are diminished for infinite in-plane bending stiffness $a_1$. In the limit of $a_1/R_c\to 0$ the motions become uncoupled as expected for the limit of a open polymer. The quality of the twist-bend coupling between transversal bending and twisting can be read of the descent of the mean square diameter depicted in Fig.~\ref{fig_diametergen}. The coupling is strongest in the regime of small $\alpha$ and $\tau$, there a small increase in $\alpha$ is compensated for by a small decrease in $\tau$, yielding constant descent. The increased transversal fluctuations are reduced by a higher inability to twist. For larger $\alpha$ the coupling to $\tau$ becomes less and less as an increase in $\alpha$ requires a larger and larger decrease in $\tau$ for compensation. The twist-bend coupling vanishes for equal bending stiffnesses $\alpha=1$ or zero twist stiffness $\tau=\infty$ as expected. 

The effective stiffening due to the ring geometry can be analyzed by comparing the amount by which the in-plane bending modes and the transversal bending modes make the diameter decrease from its mechanical equilibrium value. For an uncoupled open polymer ribbon both fluctuations are equal for equal bending stiffnesses. However, if periodic boundary conditions are applied to generate a polymer ring, in-plane fluctuations become strongly constrained while the transversal motion is only constrained via the bend-bend coupling. Discretizing a polymer ring to a polygon the restrictions on both bending motions due to the boundary conditions become obvious. While a single segment of an inextensible polygon can be moved transversal rotating by any angle up to $180^{\circ}$, a purely in-plane state is only achieved rotating by exactly $180^{\circ}$. All other in-plane bending moves would require neighboring segments to move as well, the motion is no longer truly local. In-plane bending is therefore strongly constrained, which decreases their fluctuation amplitude with respect to the transversal bending amplitude. Based on the expression for the mean square diameter given in equation (\ref{eq_d2}) the origin of the subtrahends that make the diameter decrease of the fluctuating ribbon from $D=2R_c$ and their amount are comparable. For $\tau$ and $\alpha$ small the in-plane fluctuations account for the majority of the subtrahend, as transversal fluctuations are suppressed by large twisting stiffness and large transversal bending stiffness relative to the in-plane bending stiffness. This changes if one of the relative stiffnesses increases, equal decrease of the diameter is reached along the white contour line in Fig.~\ref{fig_diametergen}. Along this line $a_2$ is well larger than the in-plane bending stiffness $a_1$, nevertheless transversal fluctuations extract more length from the diameter than in-plane modes. This puzzling behavior demonstrates the effective stiffening due to conformational constraints which mainly affects the in-plane motion. 

For reasons of illustration two limiting cases of the mean square diameter of a polymer ribbon are discussed in the following. First the limit of a double stranded bundle with $\alpha=0$ is considered. Afterwards a polymer ribbon with symmetric cross section, $\alpha=1$, is investigated. Two rigidly connected polymer strands suppress any fluctuations parallel to their interconnection, this behavior is recovered by our ribbon model in the limiting case of $a_2=\infty$. Corresponding to $\alpha=0$, the mean square diameter is easily calculated to be:
\begin{multline}
\langle D^2\rangle_{\alpha=0}=4 R_c^2-\frac{R_c^3}{ a_1}\Bigg[\frac{4+\tau\pi^2}{2\pi(1-\tau)}+\frac{2\cot(\pi\sqrt{1-\tau})}{\sqrt{1-\tau}}\\
+\frac{\tan(\pi\sqrt{1-\tau}/2)}{(1-\tau)^{3/2}}\Bigg]\;.
\label{diameterazinf}
\end{multline}  
Transversal fluctuations are still possible for any finite twist stiffness, they even increase rapidly from zero at $\tau=0$ as can be anticipated from the small distance of the contour lines for $\alpha=0$ in Fig.~\ref{fig_diametergen}. For $\tau=\infty$  the subtrahend coincides with that of $\alpha=1$. Hence, zero twist stiffness $a_3$ enables local twisting which washes out any difference between the bending stiffness $a_1$ and $a_2$, even an infinite one.

To verify our predictions based on the ribbon model by comparison with previous results and our simulations we study the special case of a symmetric cross section where both bending stiffnesses coincide $a_1=a_2=l_p$, i.e.~$\alpha=1$. This can be identified as an semiflexible polymer constrained to form a ring with persistence length $l_p$ and additional twist stiffness. The mean square diameter is computed to be: 
\begin{eqnarray}
\langle D^2\rangle_{\alpha=1}=(2R_c)^2\left(1-\frac{1}{16}\frac{2\pi R_c}{l_p}\right).
\label{eqn_alphazero}
\end{eqnarray} 
This linear decay is up to one-hundredth identical to the numerical result obtained by Shimada and Yamakawa for the mean square radius of gyration \cite{shimada}, which coincides with the mean square radius for semiflexible rings as the center of mass lies in the center of the rigid ring. Note further that the mean square diameter is independent of the twist stiffness, although no constraints where applied to the latter. In the limit of symmetric cross section the $\delta \theta$ modes become independent of $\tau$ although the $\delta\psi$ modes are still weakly coupled to bending, as can be seen by entering $\alpha=1$ into equations (\ref{eq_modetheta}) and (\ref{eq_modepsi}). Since the mean square diameter depends on $\delta\phi$ and $\delta\theta$ modes only, the mean square diameter of a ring with symmetric cross section is independent of the twist stiffness. Comparing the mean square diameter in this limiting case with the end-to-end distance of a semiflexible polymer quantifies the effective stiffness resulting from the geometry of the ring. For a weakly fluctuating semiflexible polymer of length $L$ and persistence length $l_p$ the end-to-end distance is $\langle R^2\rangle\approx L^2(1-L/3l_p)$. Examining the prefactors of the linear term we note that a ring is effectively about five times stiffer than an unconstrained polymer. This effective stiffening becomes also apparent when analyzing the shape of semiflexible polymer rings \cite{alim}. There, the stiff limit dominated by planar, elliptical shapes extends up to $L/l_p\approx 5$, beyond which three dimensional, crumpled structures prevail.
\begin{figure}[tbp]
\centering
\includegraphics[width=0.45\textwidth]{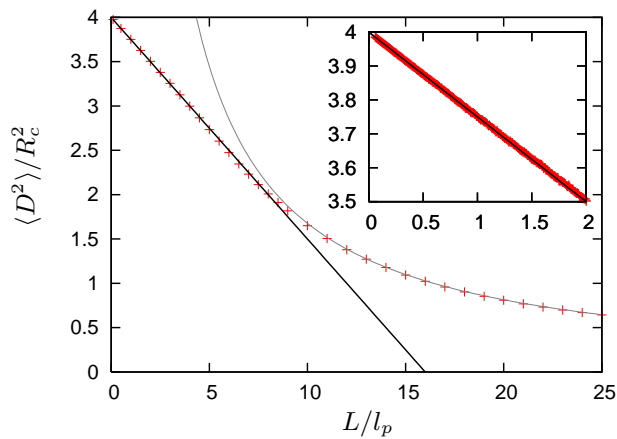}
\caption{Comparison of simulation data (crosses) for the mean square diameter $\langle D^2\rangle/R_c^2$ versus $L/l_p$ to equation (\ref{eqn_alphazero}) (black line). For $L/l_p\gg1$ the mean square diameter decays with $(L/l_p)^{-1}$ in accordance with previous results \cite{shimada} (grey line). Error bars for the Monte Carlo data are approximately of the size of the symbols.}
\label{fig_diameter}
\end{figure}

To asses the quality of the tight ring approximation a semiflexible ring with symmetric cross section and zero twist stiffness has been simulated using Metropolis Monte Carlo methods. The ring is described as a polygon composed of $N$ tethers of fixed length $a=(L/\pi)\sin(\pi/N)$ and direction $\mathbf{t}_3^i$. The energy assigned to an individual configuration is given by the elastic energy, $F=Nk_BT(l_p/L)\times$ $\times\sum_{i=1}^{N}(1-\mathbf{t}_3^i\mathbf{t}_3^{i+1})$, imposing periodic boundary conditions , $\mathbf{t}_{N+1}=\mathbf{t}_1$. New conformations are achieved by pivot moves \cite{klenin}, performing $10^6$ Monte Carlo steps per segment. The results for the mean square diameter versus flexibility $L/l_p$ presented in Fig.~\ref{fig_diameter} suggest that our approximation is valid for $L/l_p$ up to seven, only at this flexibility the cross over to nearly flexible behavior as investigated by Shimada and Yamakawa \cite{shimada} occurs. Note that there is no parameter to adjust. Since there is almost no cross-over region, a semiflexible polymer ring is entirely described by a stiff regime and an almost flexible regime, which are now both explained analytically.\\
In general weakly bending approximations yield good predictions for flexibilities up to one. However, our model for a weakly fluctuating ring is in agreement with simulations up to much higher flexibilities as a result of the effective stiffening due to the ring geometry.
\section{Conclusions}
In summary, we have investigated the effects of a ring geometry on a thermally fluctuating polymer ribbon. Parameterizing the ribbon in terms of Euler angles and approximating for small fluctuations an analytical expression for the mean square diameter of the ring was derived. Analysis of this ensemble parameter and the elastic free energy of the fluctuating semiflexible polymer constrained to a ring reveiled coupling between in-plane and transversal bending motions as well as twisting motions. Furthermore an effective stiffening of the ring mainly affecting the in-plane bending modes was found to result from the ring geometry and could be quantified as approximately five times the polymer's bending stiffness in the case of a symmetric cross section. Comparison with Monte Carlo data shows good quantitative agreement with our model up to high degrees of flexibility ($L/l_p\approx7$). Knowledge of an analytical result for the mean square diameter depending on two bending stiffness and twist stiffness enables new possibilities for the experimental determination of polymer and polymer bundle mechanics. As a thorough understanding of polymer properties is required for the unravel of cytoskeleton mechanics, we hope that our work will contribute to investigations in the broader field of cell rheology and bundle dynamics.

\begin{acknowledgement}
 Financial support of the German Excellence Initiative via the program "Nanosystems Initiative Munich (NIM)" and of the Deutsche Forschungsgemeinschaft through SFB 486 is gratefully acknowledged.
\end{acknowledgement}


\begin{thebibliography}{10}

\bibitem{adams04}
J.~C. Adams, Current Opinion In Cell Biol. {\bf 16},  590  (2004).

\bibitem{bartles00}
J.~R. Bartles, Current Opinion In Cell Biol. {\bf 12},  72  (2000).

\bibitem{tilney89}
M.~S. Tilney, L.~G. Tilney, R.~E. Stephens, C. Merte, D. Drenckhahn, D.~A. Cotanche, A. Bretscher, J. Cell Biol. {\bf 109},  1711  (1989).

\bibitem{lin94}
C.~S. Lin, W.~Y. Shen, Z.~P. Chen, Y.~H. Tu, and P. Matsudaira, Mol. Cellular Biol. {\bf 14},  2457  (1994).

\bibitem{claessens}
M.~M. A.~E. Claessens, M. Bathe, E. Frey, and A.~R. Bausch, Nature Mat. {\bf
  5},  748  (2006).

\bibitem{kratkyporod}
O. Kratky and G. Porod, Rec. Trav. Chim. {\bf 68},  1106  (1949).

\bibitem{saito}
N. Sait\^o, K. Takahashi, and Y. Yunoki, J. Phys. Soc. Jap. {\bf 22},  219
  (1967).

\bibitem{landauelas}
L.~D. Landau and E.~M. Lifschitz, {\em Theory of Elasticity} (Pergamon Press,
  Oxford, 1970).

\bibitem{everaers}
R. Everaers, R. Bundschuh, and K. Kremer, Europhys. Lett. {\bf 29},  263
  (1995).

\bibitem{liverpool}
T.~B. Liverpool, R. Golestanian, and K. Kremer, Phys. Rev. Lett. {\bf 80},  406
  (1998).

\bibitem{golestanian}
R. Golestanian and T.~B. Liverpool, Phys. Rev. E {\bf 62},  5488
  (2000).

\bibitem{mergell}
B. Mergell, M.~R. Ejtehadi, and R. Everaers, Phys. Rev. E {\bf 66},  011903
  (2002).

\bibitem{heussinger}
C. Heussinger, M. Bathe, and E. Frey, Phys. Rev. Lett. {\bf 99},  048101
  (2007).

\bibitem{bathe}
M. Bathe, C. Heussinger, M.~M.~A.~E. Claessens, A.~R. Bausch, and E. Frey, (2007), Biophys. J., in press.

\bibitem{benhamdna}
C.~J. Benham and S.~P. Mielke, Annu. Rev. Biomed. Eng. {\bf 7},  21  (2005),
  and references therein.

\bibitem{shimada}
J. Shimada and H. Yamakawa, Biopolymers {\bf 27},  657  (1988).

\bibitem{hearst}
J.~E. Hearst and N.~G. Hunt, J. Chem. Phys. {\bf 95},  9322  (1991).

\bibitem{klenin}
K.~V. Klenin, A.~V. Vologodskii, V.~V.  Anshelevich, A.~M. Dykhne, and M.~D. Frankkamenetskii, J. Mol. Biol. {\bf 217},  413  (1991).

\bibitem{rabin01}
S. Panyukov and Y. Rabin, Phys. Rev. E {\bf 64},  011909  (2001).

\bibitem{bauer}
W.~R. Bauer, R.~A. Lund, and J.~H. White, Proc. Natl. Acad. Sci. USA {\bf 90},
  833  (1993).

\bibitem{cloutier}
T.~E. Cloutier and J. Widom, Proc. Natl. Acad. Sci. USA {\bf 102},  3645
  (2005).

\bibitem{elbaum}
M. Elbaum, D.~K. Fygenson, and A. Libchaber, Phys. Rev. Lett. {\bf 76},  4078
  (1996).

\bibitem{white}
J.~H. White, Am. J. Math. {\bf 91},  693  (1969).

\bibitem{fuller}
F.~B. Fuller, Proc. Natl. Acad. Sci. USA {\bf 68},  815  (1971).

\bibitem{alim}
K. Alim and E. Frey, Phys. Rev. Lett. {\bf 99}, 198102 (2007).

\end{thebibliography}
\end{document}